\documentclass[preprint,showpacs,preprintnumbers,amsmath,amssymb]{revtex4}

\usepackage{graphicx}
\usepackage{bm}

\begin{document}


\title{Robustness of the noise-induced phase synchronization\\ in a general class of limit cycle oscillators}
\author{Jun-nosuke Teramae}
\email{teramae@brain.inf.eng.tamagawa.ac.jp}
\affiliation{Brain Science Research Center, Tamagawa University,
Machida, Tokyo, Japan}
\author{Dan Tanaka}
\email{dan@ton.scphys.kyoto-u.ac.jp}
\affiliation{Department of Physics, Graduate School of Sciences, Kyoto University, Kyoto 606-8502, Japan}

\begin{abstract}   
 We show that a wide class of uncoupled limit cycle oscillators can be
 in-phase synchronized by common weak additive noise. An expression of
 the Lyapunov exponent is analytically derived to study the stability of
 the noise-driven synchronizing state. The result shows that such a
 synchronization can be achieved in a broad class of oscillators with
 little constraint on their intrinsic property. On the other hand, the
 leaky integrate-and-fire neuron oscillators do not belong to this
 class, generating intermittent phase slips according to a power low
 distribution of their intervals. 
\end{abstract}


\maketitle

Populations of nonlinear oscillators can be found in a variety of
phenomena, including laser array \cite{wang88}, semiconductors 
\cite{hadley88wiesenfeld96}, chemical
reactions \cite{kuramoto84}, society of living organisms 
\cite{winfree67winfree80}, and neurons \cite{gray89sompolinsky91}. 
In many of these systems, the phases of oscillations can
precisely coincide owing to mutual interactions among
oscillators. Alternatively, a strong periodic input may synchronize
independent oscillators through the entrainment to the common input. In
either cases, external or internal noise sources may disturb the phase
synchronization, and therefore have long been considered to exert a
negative influence on the precise temporal relationship between
oscillators.

This view, however, has been challenged recently
\cite{pikovsky84,pikovsky01}. Pikovsky studied, in his pioneering work, a
population of circle
maps stimulated by impulse inputs at discrete random times, and found
that the common noise can induce stable phase synchronization. Since the
noise-driven synchronization does not depend on the intrinsic frequency
of oscillators, it differs from the entrainment to an external periodic
input. Evidence is accumulating for the common noise-induced
synchronization in several biological and physical systems. For
instance, an ensemble of independent neuronal oscillators may be
synchronized by a fluctuating input applied commonly to all of
them. This is suggested by experimental studies of neural information
coding \cite{mainen95}, in which the reproducibility of spike firing was tested
for a repeated application of the same fluctuating or constant input
current. Interestingly, the reproducibility of the output spike trains
was much higher for the fluctuating input than for the constant one
\cite{mainen95,tang97,jensen98ritt03kosmidis03}, indicating a high
temporal precision of the spike responses to
noisy input. In ecological systems, common environmental fluctuations
such as climate changes may synchronize different populations separated
by a large geographical distance \cite{royama92grenfell98}. In fluid
dynamics, a common turbulent
flow may generate a synchronized motion of floating particles \cite{yu90}.

All these findings indicate an active role of noise in
synchronization of non-interacting dynamical elements. It remains,
however, unclear whether the noise induced phase synchronization is
specific to a limited class of oscillators, or can be generalized to a
broad class of oscillators. In this study of a general class of
limit-cycle oscillators, we show that common additive noise, even if it
is weak, can induce phase synchronization regardless of their intrinsic
properties and the initial conditions. Using the phase reduction method
which is applicable to an arbitrary oscillator \cite{kuramoto84}, we
analytically
calculate the Lyapunov exponent of the synchronizing state and prove
that the exponent is non-positive as long as the phase-dependent
sensitivity is differentiable up to the second order. In addition, we
investigate the scaling laws that appear in the dynamics of the relative
phase when the perfect phase synchronization is deteriorated by a
discontinuous phase-dependent sensitivity or oscillator-specific noise
sources. 

Population of $N$ identical nonlinear oscillators driven by common
additive noise are described as 
\begin{equation}
  \dot{\bm{X}_i}=\bm{F}(\bm{X}_i)+\bm{\xi}(t),
 \label{eq:1}
\end{equation}
where $i=1,\cdots ,N$ and $\bm{\xi}(t)$ is a vector of Gaussian white
noise. The elements of 
the vector are normalized as $<\xi_l (t)>=0$ and
$<\xi_l (t) \xi_m (s)>=2D_{lm}\delta(t-s)$, where $\bm{D}=(D_{lm})$ is a
variance matrix 
of the noise components. Because all the oscillators are identical and
do not interact with one another, we can study the phase synchronization
of the entire population in a reduced system of two
oscillators. Regarding the common noise as a weak perturbation to the
deterministic oscillators, the phase reduction method gives the
following dynamical equations of the phases: 
\begin{equation}
  \dot{\phi_i}=\omega+\bm{Z}(\phi_i)\cdot\bm{\xi},
 \label{eq:2}
\end{equation}
where $\omega$ is an intrinsic frequency of the unperturbed
oscillators. $\bm{Z}$ is
the phase-dependent sensitivity defined as $\bm{Z}(\phi)={\rm grad}_{\bm
X} \phi |_{\bm{X}=\bm{X}_0(\phi)}$, where $\bm{X}_0(\phi)$ is the 
unperturbed limit cycle solution determined by $\bm{F}(\bm{X})$. We
assume that $\bm{Z}$ is
differentiable at least to the second order, although $\bm{Z}$ can be
discontinuous for such oscillators that have discontinuous periodic
solutions (e.g., integrate-and-fire neurons). As we will see later, the
discontinuity of $\bm{Z}$ can significantly affect the noise-driven
synchronization. To ensure the validity of the phase reduction, the weak
noise must satisfy the condition $\omega \gg |D_{lm}|$.

Equation (\ref{eq:2}) implies that the synchronizing solution described
as $\phi_1(t)=\phi_2(t)$ is 
absorbing, i.e., once two oscillators synchronize, they always remain
synchronizing. Since the area of the phase space is limited 
($0\leq \phi_1,\phi_2 < 2\pi$), the
phase variables starting from arbitrary initial phases can reach a
neighborhood of the synchronizing solution with a finite probability in
a finite time. To prove that the synchronizing solution is stable
against perturbations, we analytically calculate the Lyapunov exponent
$\lambda$
of the solution. We note that the stochastic equation (\ref{eq:2}) should be
interpreted as a Stratonovich differential equation, since the phase
reduction method assumes the conventional variable translations in
differential equations. To remove the correlation between $\phi$ and
$\bm{\xi}$, we
translate Eq.~(\ref{eq:2}) into an equivalent Ito differential equation
\cite{stratonovich63horshemke84}: 
\begin{equation}
  \dot{\phi_i}
   =\omega+\bm{Z}'(\phi_i)\bm{D}\bm{Z}(\phi_i)
   +\bm{Z}(\phi_i)\cdot\bm{\xi},
 \label{eq:3}
\end{equation}
where dash denotes differentiation with respect to $\phi$. Suppose that the
two phases have an infinitesimally small difference $\psi=\phi_2-\phi_1$. Then,
linearization of Eq.~(\ref{eq:3}) with respect to $\psi$ gives
\begin{equation}
 \dot{\psi}
  =[{\bm (}\bm{Z}'(\phi)\bm{D}\bm{Z}(\phi){\bm )}'+\bm{Z}'(\phi)\cdot\bm{\xi}]\psi,
  \label{eq:4}
\end{equation}
where $\phi$ obeys Eq.~(\ref{eq:3}). By introducing a new variable 
$y=\log(\psi)$, Eq.~(\ref{eq:4}) is
further rewritten as
\begin{equation}
 \dot{y}
  =(\bm{Z}'\bm{D}\bm{Z})'-(\bm{Z}'\bm{D}\bm{Z}')+\bm{Z}'\cdot\bm{\xi}.
  \label{eq:5}
\end{equation}
Since the Lyapunov exponent is defined as $\lim_{T\to
\infty}(y(T)-y(0))/T$, the long time average of
right hand side of Eq.~(\ref{eq:5}) coincides with $\lambda$. Replacing
the long time 
average with the ensemble average with respect to $\bm{\xi}$, we can
represent $\lambda$ as
\begin{eqnarray}
 \lambda
  &=& <(\bm{Z}'\bm{D}\bm{Z})'-(\bm{Z}'\bm{D}\bm{Z}')+\bm{Z}'\cdot\bm{\xi}>_\xi\nonumber\\
 &=& <(\bm{Z}'\bm{D}\bm{Z})'-(\bm{Z}'\bm{D}\bm{Z}')>_\xi\nonumber\\
 &=& \int_{0}^{2\pi}P_{st}(\phi)[(\bm{Z}'\bm{D}\bm{Z})'-(\bm{Z}'\bm{D}\bm{Z}')]d\phi.
  \label{eq:6}
\end{eqnarray}
Here, the second line follows from $<\bm{Z}'(\phi)\cdot\bm{\xi}>_\xi=0$, which
holds in Ito stochastic 
processes. $P_{st}$ is a steady distribution function of $\phi$ described as 
\begin{equation}
 P_{st}(\phi)=\frac{C}{Z^2(\phi)}\int_{\phi}^{\phi+2\pi}\exp{(V(x)-V(\phi))}dx,
  \label{eq:7}
\end{equation}
with an effective potential $V(\phi)=-\int^{\phi}\frac{\omega+\bm{Z}'(x)\bm{D}\bm{Z}(x)}{\bm{Z}(x)\bm{D}\bm{Z}(x)}dx$ and a normalization constant $C$. Fortunately under the assumption of
weak noise, $\omega\gg |D_{lm}|$, $P_{st}$ is reduced to a constant
function $P_{st}(\phi)=1/(2\pi)$. By substituting
this into the last line of Eq.~(\ref{eq:6}), and noting that the first term
vanishes due to the periodicity of $\bm{Z}$, we finally obtain the following
formula: 
\begin{equation}
 \lambda=-\frac{1}{2\pi}\int_{0}^{2\pi}\bm{Z}'\bm{D}\bm{Z}'d\phi\leq 0,
  \label{eq:9}
\end{equation}
where the equality holds if $\bm{Z}$ is a constant function.
Since the variance matrix is positive definite, $\lambda$ is non-positive. This
implies that the phase synchronization induced by common noise is stable
in an arbitrary oscillator system regardless of the detailed oscillatory
dynamics, as long as $\bm{Z}$ is differentiable.

\begin{figure}
 \includegraphics{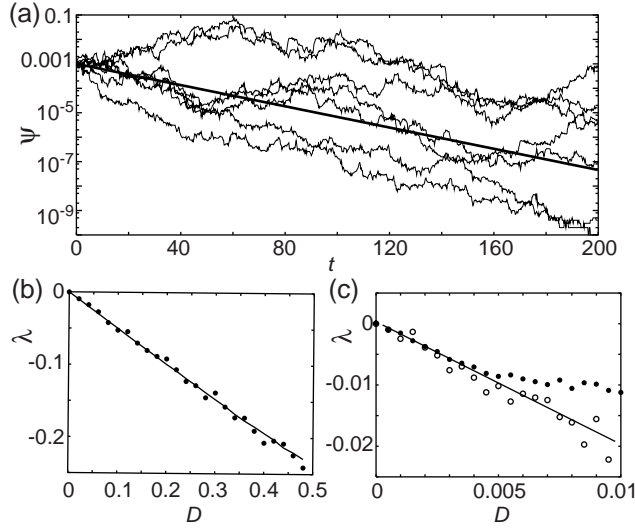}
  \caption{
 The common noise-induced phase synchronization. (a) The phase difference
 between two oscillators shows an exponential decay fluctuating around
 the theoretical behavior (solid line) derived from
 Eq.(\ref{eq:9}). Different curves
 correspond to different realizations of the random driving force
 $\xi(t)$. Here, $Z(\phi)=\sin{\phi}$, $\omega=1$ and $D=0.1$. (b) The
 Lyapunov exponent of the above 
 synchronizing solution is shown as a function of the common-noise
 intensity (circles). The solid line shows an analytical result. (c) The
 Lyapunov exponent is calculated for the synchronizing solution of the
 Stuart-Landau oscillator. Numerical results are shown for both original
 (filled circles) and reduced (empty circles) oscillator systems. The
 solid line represents a theoretical result. The parameter values are set
 as $c_0=2$, $c_2=1$, $D_{11}=D_{22}=D$ and $D_{12}=D_{21}=0$.
 }
 \label{figure:1}
\end{figure}

To confirm the above analytical results, we numerically solve Eq.~(\ref{eq:2})
to obtain the Lyapunov exponent for a specific choice of $Z$, i.e.,
$Z(\phi)=\sin(\phi)$. The
phase difference between the two oscillators driven by common additive
noise shows an exponential decay, and the decay constant well agrees
with the analytical result (Fig. \ref{figure:1}a). Consistent with equation
(\ref{eq:9}), the
magnitude of the negative Lyapunov exponent increases in proportion to
the noise intensity $D$ (Fig. 1b). In order to confirm the validity of the
phase reduction method, we employ the Stuart-Landau oscillators and
compare the Lyapunov exponent derived from Eq.~(\ref{eq:9}) with that calculated
numerically in the original oscillator system (Fig. 1c). For the
Stuart-Landau oscillator described as
$\dot{A}=(1+ic_0)A-(1+ic_2)|A|^2A$, the phase sensitivity can be 
explicitly given as
$Z(\phi)=(-c_2\cos{\omega\phi}-\sin{\omega\phi},-c_2\sin{\omega\phi}+\cos{\omega\phi})/\omega$,
where $\omega=c_0-c_2$ \cite{kuramoto84}.

\begin{figure}
 \includegraphics{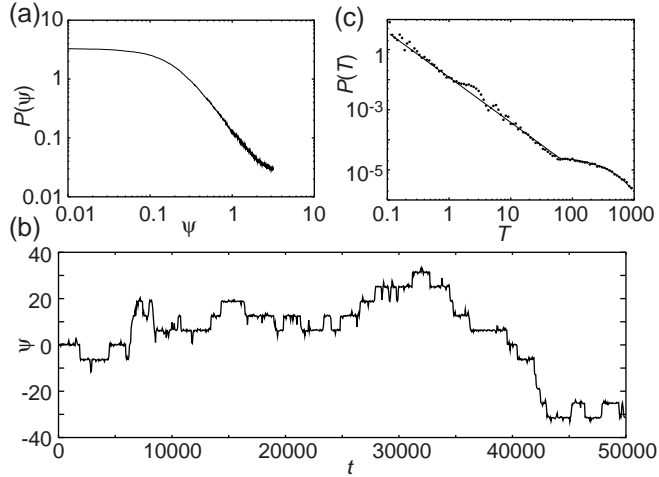}
  \caption{
 The intermittent phase slips induced by uncommon additive noise sources
 to the oscillators defined by $\omega=1$ and $Z(\phi)=\sin{\phi}$. The
 noise intensities $D=0.1$ and 
 $d=0.001$. (a) The distribution of the phase difference over a
 sufficiently long time. (b) The time evolution of the phase difference
 exhibits intermittent phase slips. (c) The distribution of the
 inter-phase-slip intervals shows a power-low decay with an exponent of
 $-3/2$ (fitted by a solid line) and an exponential cutoff resulted from
 the additive noise term in Eq.(\ref{eq:11}).
 }
 \label{figure:2}
\end{figure}

In practical situations, the individual oscillators may be influenced by
additional, oscillator-specific noise sources. They may appear from the
fluctuations intrinsic in the oscillators, or a lack of perfect
coincidences in the common driving noise. To discuss the influences of
additional noise, Eq.~(\ref{eq:1}) is modified to
\begin{equation}
  \dot{\bm{X}_i}=\bm{F}(\bm{X}_i)+\bm{\xi}(t)+\bm{\eta}_i(t),
  \label{eq:10}
\end{equation}
where the uncommon noise sources $\bm{\eta}_i$ are normalized as
$<\eta_{i,l}(t)>=0$ and
$<\eta_{i,l}(t)\eta_{j,m}(s)>=2d_{lm}\delta_{ij}\delta(t-s)$.
Linearization of Eq.~(\ref{eq:10}) gives the stochastic equation of the
phase
difference as
\begin{equation}
 \dot{\psi}
  =\left[
    {\bm (}\bm{Z}'(\bm{D}+\bm{d})\bm{Z}{\bm )}'+\bm{Z}'\cdot(\bm{\xi}+\frac{\bm{\eta}_1+\bm{\eta}_2}{2})
    \right]
    \psi+\bm{Z}\cdot(\bm{\eta}_2-\bm{\eta}_1).
  \label{eq:11}
\end{equation}
Since both multiplicative and additive factors fluctuate, Eq.~(\ref{eq:11}) is
regarded as a multiplicative stochastic process with additive noise,
which has been studied in variety of fields
\cite{kuramoto97anakao98teramae01}. The steady
distribution function of Eq.~(\ref{eq:11}) exhibits a power-low decay in
a middle
range of $\psi$ (Fig. 2a), implying that for most of the time the phases stay
near phase synchronization even in the presence of uncommon noise
sources. Simulations of Eq.~(\ref{eq:10}) reveal that the system is
trapped in the 
phase synchronizing state for certain intervals between intermittent
phase slips (Fig. 2b). It is known that the intermittent bursts are
characteristic to the stochastic processes driven simultaneously by
multiplicative and additive noise sources \cite{cenya}, and that the
intervals between neighboring bursts obey a power low distribution with
an exponent of $-3/2$. In Fig. 2c, the inter-slip intervals of the present
phase dynamics obeys a power low distribution of the same exponent.

\begin{figure}
 \includegraphics{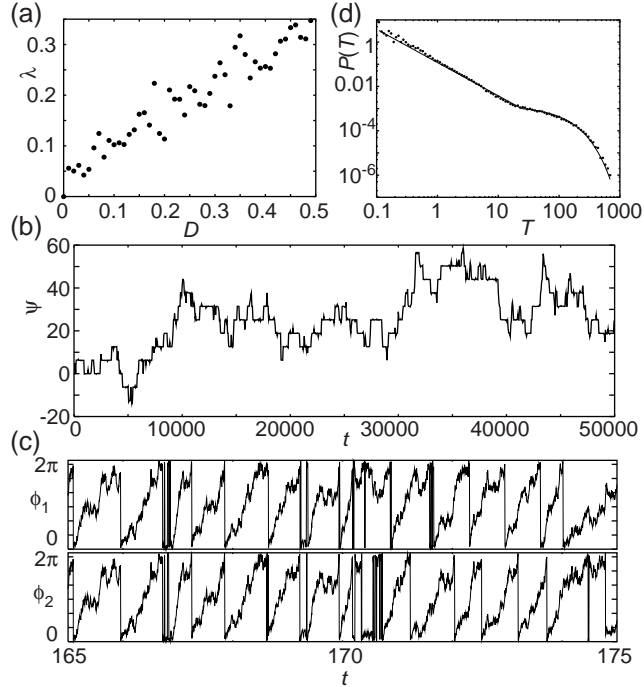}
  \caption{
 The unstable phase synchronization of integrate-and-fire models
 stimulated by a common noise source. The phase-dependent sensitivity of
 this oscillator is discontinuous, and the noise-driven
 synchronization is not ensured. $I=2$ and no uncommon noise, $d=0$. (a)
 The Lyapunov exponent $\lambda$ 
 is plotted as a function of the common noise strength $D$. (b) The phase
 difference shows significant jitters, diffusing due to intermittent
 phase slips. Here and below, $D=0.1$. (c) The time evolution of the two
 phasess displays temporarily synchronizing
 ($t<170$) and desynchronizing ($t>170$) states. (d) The distribution of the
 inter-phase-slip intervals shows an exponential decay with an exponent of
 $-3/2$ (fitted by a solid line).
 }
 \label{figure:3}
\end{figure}

So far, we have assumed that the phase-dependent sensitivity $\bm{Z}$ is a
continuous function of the phase. However, the phase reduction method
does not ensure the continuity of $\bm{Z}$, and some oscillators do not have
this property. For example, an integrate-and-fire neuron oscillator,
which is described by $\dot{v}=I-v$ with a renewal condition
$v(t)=1\to\lim_{\tau\to0+}v(t+\tau)=0$, is frequently used for
modeling neuronal activity, but it has the following discontinuous $\bm{Z}$:
\begin{equation}
 Z(\phi)
  =\frac{\omega}{I}\exp{(\frac{\phi}{\omega})},~~0\leq\phi<2\pi,
  \label{eq:12}
\end{equation}
where $\omega=2\pi/(\log{I}-\log{(I-1)})$. As shown in Fig. 3a,
numerical integrations of Eq.~(\ref{eq:2}) for the $\bm{Z}$ 
given in Eq.~(\ref{eq:12}) show positive Lyapunov exponents, with the
magnitudes 
increased with the intensity of the common noise. The phase difference
does not decay exponentially, but fluctuates around the synchronizing
state between the intermittent phase slips (Fig. 3b). Figure 3c displays
the synchronized time evolution of the phase variables that is
terminated by an abrupt phase slip at $t\approx 170$: After that, the
two phases 
are desynchronized until they recover the phase synchronization (not
shown). As in the previous case, the inter-slip intervals obey a $-3/2$
power-low distribution (Fig. 3d). This intermittency is essentially the
same as the on-off intermittency of chaotic oscillators just before the
onset of synchronization, thus associated with positive Lyapunov
exponents \cite{fujisaka85fujisaka86}. Note that the discontinuity of
the phase sensitivity 
and the resultant positive Lyapunov exponents are inherent in the leaky
integrate-and-fire model. For example, the Hodgkin-Huxley model has a
continuous and differentiable phase sensitivity, thus yielding a
negative Lyapunov exponent and a stable phase synchronization in
response to common noise (results not shown).

We briefly argue the relationships between the present study and two
previous studies. In the stochastic resonance, the ability of an
excitable system in detecting a weak signal can be optimized by noise of
suitable intensity \cite{benzi81gammaitoni98}. The present study also
argues the role of 
noise in improving the response reliability. However, here the
improvement is achieved by the precise temporal coincidences between
oscillators, whereas the stochastic resonance only enhances the response
probability without caring the exact timing of events. Thus, the
stability or Lyapunov exponent is not a central issue in the stochastic
resonance, and the two studies deal with qualitatively different
phenomena. Second, some chaotic oscillators exhibited phase
synchronization when they were driven by common additive noise
\cite{pikovsky01,zhou02}. However, it remained unknown whether this type
of synchronization may appear in a
broad class of, either chaotic or non-chaotic, oscillator systems. In
this paper, we have proven that such a
synchronization can be induced in a wide class of limit-cycle
oscillators. A unified treatment of limit cycle oscillators and chaotic
oscillators is awaited, as they may share many characteristic properties
of the common noise-induced synchronization.

In conclusion, independent limit cycle oscillators can be synchronized
by weak, common additive noise regardless of the detailed oscillatory
dynamics and the initial phase distributions. The stability of this
synchronizing solution only requires the presence of a second derivative
of the phase-dependent sensitivity, so the solution can exist in a broad
class of oscillators. The leaky integrate-and-fire oscillators do not
belong to this class of oscillators, and show no perfect
synchronization. The analytical evaluation of the Lyapunov exponent
remains open for further studies for oscillators possessing
non-differentiable phase sensitivity.

The authors are very grateful to T. Fukai for a critical reading of the
manuscript and helpful comments. The authors thank T. Aoyagi,
H. Nakao and Y. Tsubo for discussions. J.T. was supported
by the 21st
Century COE Program of Japanese Ministry of Education, Culture, Sprots,
Science and Technology. D.T. acknowledges gratefully 
financial support by the Japan Society for Promotion of Science (JSPS).
This work was partially supported by Grant-in-Aid for Scientific
Research (B)(2) 16300096.


\begin{thebibliography}{99}
\bibitem{wang88}
S. S. Wang and H. G. Winful,
Appl. Phys. Lett. {\bf 52}, 1774 (1988).

\bibitem{hadley88wiesenfeld96}
P. Hadley, M. R. Beasley and K. Wiesenfeld,
Phys. Rev. B {\bf 38}, 8712 (1988);
K. Wiesenfeld, P. Colet and S. H. Strogatz, 
Phys. Rev. Lett. {\bf 76}, 404 (1996).

\bibitem{kuramoto84}
Y. Kuramoto,
{\em Chemical Oscillation, Waves, and Turbulence}
(Springer-Verlag, Tokyo, 1984); (Dover Edition, 2003).

\bibitem{winfree67winfree80}
A. T. Winfree, 
J. Theor. Biol. {\bf 16} (1967), 15;
A. T. Winfree,
{\em The Geometry of Biological Time}
(Springer, New York, 1980).

\bibitem{gray89sompolinsky91}
C. M. Gray, P. K{\" o}nig, A. K. Engel and W. Singer,
Nature {\bf 338}, 334 (1989);
H. Sompolinsky, D. Golomb and D. Kleinfeld,
Phys. Rev. A {\bf 43}, 6990 (1991).

\bibitem{pikovsky84}
A. S. Pikovsky,
In R. Z. Sagdeev, Editor, {\em Nonlinear and Turbulent Processes in
	Physics}, 1601, (Harwood, Singapore, 1984).

\bibitem{pikovsky01}
A. S. Pikovsky, M. Rosenblum, and J. Kurths,
{\em Synchronization -A Unified Approach to Nonlinear Science}
(Cambridge University Press, Chambridge, U.K., 2001).

\bibitem{mainen95}
Z. F. Mainen and T. J. Sejnowski
Sience {\bf 268}, 1503 (1995).

\bibitem{tang97}
A. C. Tang, A. M. Bartels and T. J. Sejnowski,
Cerebral Cortex {\bf 7}, 502 (1997).

\bibitem{jensen98ritt03kosmidis03}
R. V. Jensen,
Phys. Rev. E {\bf 58}, R6907 (1998);
J. Ritt,
Phys. Rev. E {\bf 68}, 041915 (2003);
E. K. Kosmidis and K. Pakdaman,
J. Comput. Neurosci. {\bf 14}, 5 (2003).

\bibitem{royama92grenfell98}
T. Royama,
{\em Analycal Population Dynamics}
(Chapman and Hall, London, 1992);
B. T. Brenfell et al.,
Nature, {\bf 394}, 674 (1998).

\bibitem{yu90}
L. Yu, E. Ott, and Q. Chen,
Phys. Rev. Lett. {\bf 65}, 2935 (1990).

\bibitem{stratonovich63horshemke84}
R. L. Stratonovich,
{\em Topics in the Theory of Random Noise}
(Gordon and Breach, New York, 1963);
W. Horshemke and R. Lefever,
{\em Noise-Induced Transitions}
(Springer-Verlag, Berlin, 1984).

\bibitem{kuramoto97anakao98teramae01}
Y. Kuramoto and H. Nakao,
Phys. Rev. Lett. {\bf 78}, 4039 (1997);
H. Nakao,
Phys. Rev. E {\bf 58}, 1591 (1998);
J. Teramae and Y. Kuramoto,
Phys. Rev. E {\bf 63}, 036210 (2001).

\bibitem{cenya}
A. {\v C}enya, A. N. Angnostopoulos, and G. L. Bleris,
Pys. Lett. A {\bf 224}, 346 (1997).

\bibitem{fujisaka85fujisaka86}
H. Fujisaka and H. Yamada, 
Prog. Theor. Phys. {\bf 74}, 918 (1985);
H. Fujisaka and H. Yamada, 
Prog. Theor. Phys. {\bf 75}, 1087 (1986).

\bibitem{benzi81gammaitoni98}
R. Benzi, A. Sutera and A. Vulpiani,
J. Phys. A {\bf 14}, L453 (1981);
L. Gammaitoni, P. H{\"a}nggi, P. Jung and F. Marchesoni,
Rev. Mod. Phys. {\bf 70}, 223 (1998);
T. Fukai,
Neuroreport {\bf 11}, 3457 (2000).

\bibitem{zhou02}
C. Zhou, J. Kurths,
Phys. Rev. Lett. {\bf 88}, 230602 (2002).

\end{thebibliography}
\end{document}